\documentclass[preprint,preprintnumbers,amsmath,amssymb,showpacs,superscriptaddress]{revtex4}
\usepackage{graphicx}
\usepackage{dcolumn}
\usepackage{bm}
\usepackage[usenames]{color}

\begin{document}

\title{{\bf Medium Effects in DIS from Polarized Nuclear Targets}}

\author{H. Fanchiotti}
\affiliation{Departamento de F\'isica, Universidad Nacional de La
Plata,  C.C. 67-1900 La Plata, Argentina} \affiliation{IFLP
(CONICET), Universidad Nacional de La Plata,  C.C. 67-1900 La
Plata, Argentina}
\author{C. A. Garc\'ia Canal}
\affiliation{Departamento de F\'isica, Universidad Nacional de La
Plata,  C.C. 67-1900 La Plata, Argentina}
\affiliation{IFLP
(CONICET), Universidad Nacional de La Plata,  C.C. 67-1900 La
Plata, Argentina}
\author{T. Tarutina}
\affiliation{Departamento de F\'isica, Universidad Nacional de La
Plata,  C.C. 67-1900 La Plata, Argentina} \affiliation{IFLP
(CONICET), Universidad Nacional de La Plata,  C.C. 67-1900 La
Plata, Argentina}
\author{V. Vento}
\affiliation{Departamento de F\'{\i}sica Te\'{o}rica and Instituto de F\'{\i}sica Corpuscular, Universidad de
Valencia\\ Consejo Superior de Investigaciones Cient\'{\i}ficas, 46100 Burjassot (Val\'{e}ncia),
Spain}

\begin{abstract}
The behavior of the nucleon structure functions in lepton nuclei deep inelastic scattering, both polarized and unpolarized, due to nuclear structure effects is reanalyzed. The study is performed in two schemes: an {\it{x}}-rescaling approach, and one in which there is an increase of sea quark components in the in medium nucleon, related to the low energy {\it{N-N}} interaction. In view of a recent interesting experimental proposal to study the behavior of the proton spin structure functions in nuclei we proceed to compare these approaches in an effort to  enlighten the possible phenomenological interest of such difficult experiment.
\end{abstract}

\maketitle
\section{Introduction}
More than 30 years ago, the European Muon Collaboration (EMC) discovered that the unpolarized structure function of a bound nucleon in a nucleus is different from that of a free nucleon, and different from one nucleus to another \cite{Aubert:1983xm}. This experimental fact triggered an innumerable series of analyses and their corresponding explanations (see \cite{Rith:2014tma} for an extensive list of references) and very interesting recent theoretical developments relating the EMC effect to Short Range Correlations \cite{Higinbotham:2010tb,Weinstein:2010rt,Piasetzky:2011zz,Arrington:2012ax,Hen:2012fm} ,\cite{Frankfurt:2009vv,Vanhalst:2012ur,GarciaCanal:2013dma}.

Not long ago there has been a proposal for an experimental study of the nuclear effect for the polarized structure function $g_1$, an experiment in which both projectile and the nuclear target are longitudinally polarized \cite{Joo}. This work has been motivated by some detailed model dependent calculations \cite{Cloet:2005rt,Cloet:2006bq,Smith:2005ra,Ganesamurthy:2011zza}, which have coined the development: {\it polarized EMC effect}. This problem was theoretically studied long ago   for the Deuteron by Frankfurt and Strikman \cite{Frankfurt:1981mk} and shortly thereafter for arbitrary spin by Jaffe and Manohar \cite{Jaffe:1988up} with a non-relativistic convolution model. The recent analyses are detailed model calculations which take into account a realistic low energy nuclear description, a model for hadron structure and include QCD evolution in their schemes. Their result shows a large effect in the nuclear to proton ratio and moreover,  the contribution of the quark could dramatically affect  it at small $x$.

Despite the fact that the proposed phenomenon is very interesting and might help to understand the behavior of the nucleons in nuclei,  the name, as pointed out by the PAC29 committee \cite{PAC29} might be misleading. In the experimental proposal the chosen nucleus has been  $^7 $Li, were the spin of the system, $3/2$, is naively mostly associated with a valence proton lying in a $p$-shell, while the remaining nucleons are coupled mostly in pairs to total angular momentum $0$. Thus the averaged medium behavior implied by the unpolarized EMC effect is not present in this polarized case. However, it is clear that the active valence proton is subject to the effect of its companions and therefore a medium effect must be present.

Our aim in this note is, by avoiding in as much as possible model dependence, to center our attention into physical ideas which might be at the origin of the discussed phenomenon. For that purpose we revisit the unpolarized EMC effect in terms of an $x$-rescaling description \cite{GarciaCanal:1984eh} and  a pion content approach  \cite{Epele:1994aq,deFlorian:1993hx} to fix the ideas and the parameters connected with the experimental data available at present. We then proceed to study the in medium polarized case.  We  have in mind in all our discussion the experimentally proposed $^7$ Li, which represents an ideal system to distinguish conceptually between the { \it polarized} and unpolarized EMC effects. We end by comparing the polarized and unpolarized phenomena with the intention of motivating further experimental research.

\section{Unpolarized DIS}

In this section we briefly summarize the analysis of DIS with nuclear targets in the two approaches mentioned before, the  $x$-rescaling approach \cite{GarciaCanal:1984eh},  and a qualitative approximation to the so called pion content approach \cite{Epele:1994aq,deFlorian:1993hx}.

This $x$-rescaling approach contains only one parameter, $\eta$,  and the EMC effect was described by suggesting that
the true scaling variable for deep-inelastic scattering off
nuclei should be taken to be $x^* = \eta \,x$ \cite{GarciaCanal:1984eh}. The main idea behind this approach  is that the  quark distributions in nuclei are shifted towards lower $x$ values as compared to those corresponding to free nucleons. Thus the name for the mechanism, $x$-rescaling.  This approach  was shown to be connected \cite{GarciaCanal:1986xe}  to the $Q^2$-rescaling approach \cite{Close:1984zn}.

 Consequently, the measured ratios of the nuclear to deuteron
structure functions can be written, for fixed $Q^2$  as,
\begin{equation}
R(A) = F_2^{A}(x^*)/F_2^{D}(x^*)
\label{f2A}
\end{equation}
where $F_2^{A}(x^*)$ is the nuclear structure function calculated
using a rescaled variable $x^*$ in the free proton and neutron
structure functions, and $F_2^{D}(x^*)$ is the Deuteron structure
function where the effects of rescaling are small.

The $x$-rescaling mechanism leads to fits of very good quality
for the EMC effect in the region $0.3< x < 0.75$ for all the nuclei experimentally analyzed \cite{GarciaCanal:1984eh,GarciaCanal:2013dma}. Recently, this approach was used to show an interplay between the quark-gluon and hadronic degrees of freedom in the unpolarized EMC \cite{GarciaCanal:2013dma}.

The second approach is a simplification of the so called pion content model \cite{Epele:1994aq,deFlorian:1993hx}.  In this  approach one introduces the pion presence into the nucleon structure function and makes a convolution model which takes this presence and the pion structure into account. The end result is twofold, on the one hand there is a $x$-rescaling effect associated to momentum conservation in the pion emission and on the other hand there is an explicit contribution of the sea associated with the pionic structure function.  This approach also leads to fits of very good quality for the EMC effect in the whole $x$-region for all the nuclei experimentally analyzed. The direct contribution of the sea is  important for low $x$ but not so much for intermediate $x$. However, and this is a peculiarity of this approach, it is necessary to consider also the pion sea effect  in the deuteron when one deals with the EMC ratio, if one wants to avoid the blowing up of the ratio at  low $x$ which  would destroy the agreement in the EMC ratio \cite{Epele:1994aq,deFlorian:1993hx}.

We  incorporate this twofold mechanism, in our phenomenological scheme, by means of two parameters. One, $\delta$, describes  the associated scaling mechanism into the free nucleon structure functions as before  $ x \rightarrow \delta x $;  analogously  to $\eta$, $\delta >1$.  Another,  $\sigma$, which magnifies the  contribution  from the sea associated to the $u$ and $d$ flavors in the intermediate $x$ region, concentrated initially for very  small $x$. For one nucleon in the medium the contribution to $F_2$ is
\begin{equation}
F_2^{N} (x) = F_2^{N} (\delta x) + F_2^{udsea_N} (\sigma x).
\end{equation}
This modified nucleon structure functions have to be incorporated into Eq. (\ref{f2A}) to perform the corresponding EMC average. We shall call this approach modified sea scheme (MSS).


\begin{figure}[htb]
\begin{center}
\includegraphics[scale= 1.]{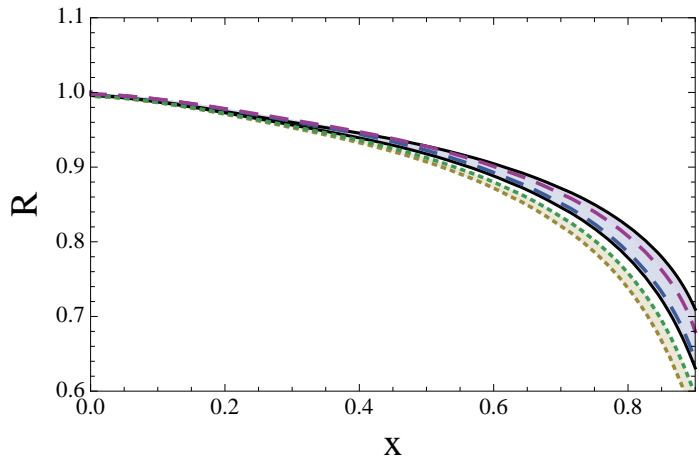}
\caption{$R$ ratio as a function of $x$ in  the $x$-scaling approach (black), in the modified sea  approach with (dashed) and without (dotted) the added sea contribution. The parameter values: for $^7$Li: $\eta= 1.011\pm 0.002,\, \delta= 1.016\pm 0.001,\, \sigma= 1$; for $D$: $\eta= 1.0, \,\delta= 1.005,\, \sigma= 1$.}
\label{f2Li}
\end{center}
\end{figure}

Our study is concentrated on the experimentally wishful nucleus $^7$Li \cite{Joo}. We use the proton and neutron structure functions and sea distributions from the analysis of Ref. \cite{Martin:2009iq}  for fixed $Q^2 = 10 $GeV$^2$. The value of $\eta$ for $^7$Li is extracted from a linear extrapolation of a fit to the data of several nuclei. The value obtained for $^7$ Li is $\eta = 1.011 \pm 0.002$. The values of the parameters for the MSS scheme are obtained by fitting them to reproduce the Ratio, $R$, of the $x$-scaling description in the EMC region and are shown in the caption of Fig. \ref{f2Li}.

\begin{figure}[htb]
\begin{center}
\includegraphics[scale= 1.]{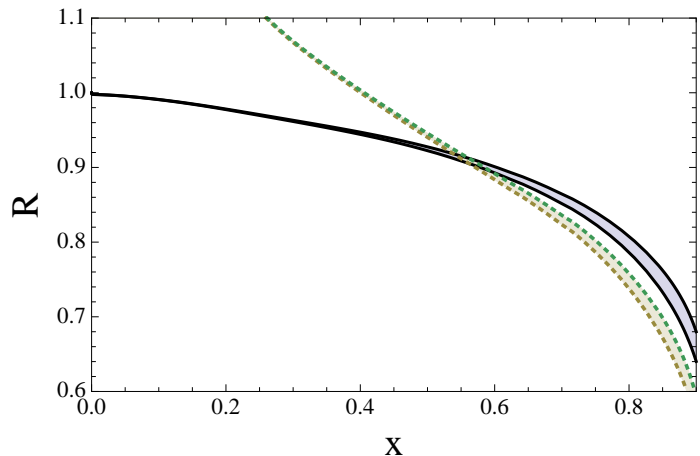}
\caption{$R$ ratio as a function of $x$ in the modified sea approach  with  (black) and without (dashed) Deuteron sea. Parameter values: for $^7$ Li: $\delta= 1.016 \pm 0.001,\, \sigma= 1$, for $D$: $\eta=1.00$}
\label{f2noDsea}
\end{center}
\end{figure}

In Fig.\ref{f2Li} we show our prediction of both approaches and the size of the sea contribution in the MSS method. A good agreement between both approaches is obtained by simply doubling the sea contribution. Note, as it was already mentioned, that we had to incorporate an extra sea contribution for the Deuteron in order to avoid a dramatic increase at the origin, where it tends to dominate. Fig. \ref{f2noDsea} shows that effect.


\section{Polarized DIS}

Several studies of the nuclear effects for the $g_1$ polarized structure functions in the typical EMC $x$ region have recently appeared. They are based on detailed dynamical nucleon structure models and models for nuclear matter. The analysis of Ref. \cite{Cloet:2005rt,Cloet:2006bq} predicts a strong increase of the ratio $R_{pol}= g_1^A/A g_1^p$ of the order of twice the size in the unpolarized case. This calculation is based upon a convolution where the nucleon is described as a bound state of a quark-diquark in the Nambu-Jona-Lasinio (NJL) model and the presence of the nuclear medium is taken into account through mean fields which act on the quarks in the nucleon. This is mainly a valence quark picture where the sea appears by evolution.  Another proposal \cite{Smith:2005ra} includes  sea quarks explicitly in order to be consistent  with both DIS and Drell-Yan results.  Consequently, the nuclear medium effects are present both in the valence and the  sea quark distributions. They are computed using the chiral quark-soliton model and the nuclear effects related to the valence quarks are similar to those obtained in Ref. \cite{Cloet:2005rt,Cloet:2006bq}, while the inclusion of sea quarks gives rise to an important increase for $x < 0.3$. Both description contain many parameters that have to be fitted.  A  third analysis \cite{Ganesamurthy:2011zza} is based on the phenomenological Thermodynamical Bag Model, a modification of the MIT bag model, and shows not much difference between the polarized and the unpolarized ratios. 

For our analysis we use the fits to all polarized data of Ref. \cite{deFlorian:2008mr,deFlorian:2009vb} for $Q^2 = 10 $GeV$^2$.  We start by the rescaling approach and as in the case  of the unpolarized structure function we replace the scaling $x$ variable by a rescaled variable $x^*$ in the free nucleon structure functions.

In order to find reasonable values for this $\eta$ parameter, we follow two arguments. The first, which we already discussed previously, is the definition of the polarized in medium effect. Let us for clarity limit ourselves to $^7 Li$. In this case,  from a naive nuclear shell model point of view  we have one active proton in a $p$-shell and the remaining nucleons are coupled to total angular momentum zero in $s$ and $p$ shells. More sophisticated  calculations with Green Function Monte Carlo \cite{Pieper:2004qw} and a Cluster Model \cite{Walliser:1985zz}  give a mean proton polarization of $87\%$  and $3\% $ for the neutron. For simplicity we assume the shell model picture and we define the ratio

\begin{equation}
R_{pol}(x) = \frac{g_1^p(x^*)}{g_1^p (x)}.
\label{g1x}
\end{equation}
It is interesting to compare now with Eq. \ref{f2A} to realize the difference between the polarized and the unpolarized effects, namely the conventional EMC averages over nucleons, while not so the polarized in medium effect.

\begin{figure}[htb]
\begin{center}
\includegraphics[scale= 1.]{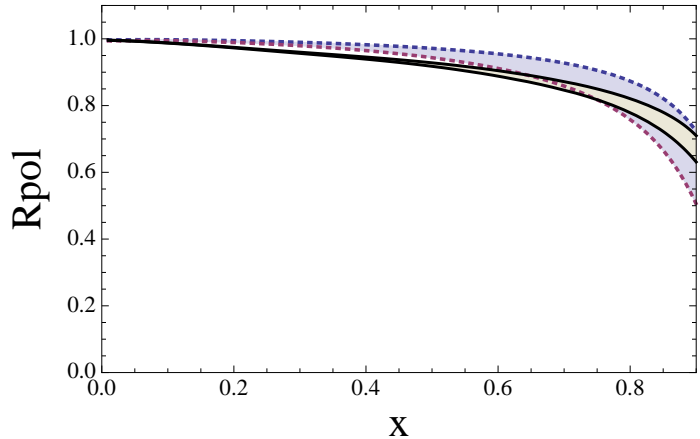}
\caption{$R_{pol}$ ratio for $^7$Li as a function of $x$ in the $x$-scaling approach. The dotted curves correspond to parameter values for $^7$Li: upper $\eta=1.01$, lower $\eta=1.02$. The black band corresponds to $R$ with $\eta = 1.011\pm 0.002$.}
\label{g1f2x}
\end{center}
\end{figure}

Accepting the previous naive definition, the second problem  we face is the relation of the $x$-scaling parameter $\eta$ with the dynamics. This parameter is related to the efective nucleon mass in the medium \cite{GarciaCanal:1984eh} which is an average property and therefore we expect that $\eta$ should not change. However, the fact that the nucleon is polarized might incorporate some spin interactions, that average out in the unpolarized case, which might modify it. The results of  Refs. \cite{Cloet:2005rt,Cloet:2006bq,Smith:2005ra}  seem to indicate that $\eta$ should increase. We plot in Fig. \ref{g1f2x} the value of the ratio Eq. (\ref{g1x}) in form of a band, whose smallest value corresponds to the unpolarized $\eta$ value and the largest to $\eta = 1.02$. The full curve shows the ratio $R$. We note that the $R_{pol}$ ratio is extremely sensitive to $\eta$ and therefore to the detailed dynamics of the polarized case.


%

\begin{figure}[htb]
\begin{center}
\includegraphics[scale= 1.]{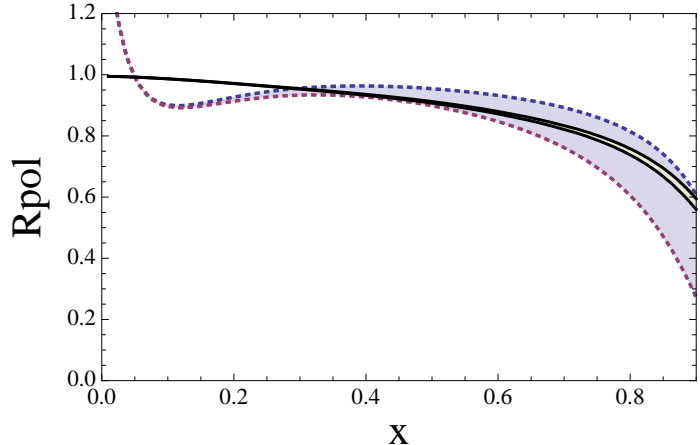}
\caption{$R_{pol}$ for $^7$Li as a function of $x$ in the MSS approach. The dotted curves correspond to parameter values for Li: upper $\delta=1.015, \,\sigma=1$, lower $\delta=1.035,\, \sigma=1.02$. The black band corresponds to $R$ with $\eta=1.011 \pm 0.002, \, \sigma = 1.0$.}
\label{g1f2sea}
\end{center}
\end{figure}

We next proceed to the MSS approach. As before the nucleus structure function will be defined in terms of modified nucleon structure functions,
\begin{equation}
g_1^N (x) = g_1^N (\delta x) + g_1^{udsea_N} (\sigma x).
\label{g1sea}
\end{equation}
For $^7 Li$, $N$ is a proton. In Fig.\ref{g1f2sea} we plot $ R_{pol}$ with the same parameters as for $R$, and following the same discussion as before we construct a band with a $2\%$ increase both in $\delta$ and  $\sigma$.


%

\begin{figure}[htb]
\begin{center}
\includegraphics[scale= 1.]{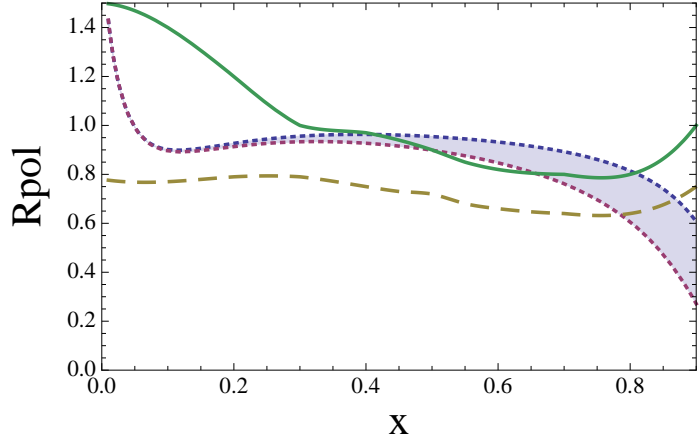}
\caption{$R_{pol}$ band for $^7$Li as a function of $x$ in the MSS approach as in the previous figure (dotted curves). The solid curve corresponds to the calculation in Ref. \cite{Smith:2005ra} with the sea contribution included, while the dashed curve the that of Ref. \cite{Cloet:2005rt,Cloet:2006bq}.}
\label{g1comparison}
\end{center}
\end{figure}

Finally in Fig.\ref{g1comparison} we compare our results with the two most extreme published analyses. Our result is intermediate between both. We could obtain a higher sea contribution close to the origin by increasing $\sigma$ in  line of Ref.\cite{Smith:2005ra}, but we can never get a result as low as that of Ref.\cite{Cloet:2005rt,Cloet:2006bq} for low $x$. 

 We have only analyzed in here the low flavor sea contributions not worrying about heavy flavors and gluons, whose contribution are small in  the two approches.

\section{Conclusion}

The schemes presented in this note parametrize the change of parton distributions in medium in terms of one or two parameters. These parametrizations are motivated by QCD based dynamical pictures, $x$-scaling associated to a Renormalization Group Analysis, and the MSS,  a  chiral scheme convoluted into a partonic description. It is really surprising that in the unpolarized case,  with one or two parameters, one is able to fit the  EMC effect for all studied nuclei.  Thus our fits lead, based on QCD and chiral dynamics, to  parametrizations of the data with a minimal number of parameters. 

What can a polarization experiment contribute to our undertanding of proton structure? As has been widely discussed this is not an EMC {``}average" type experimental analysis. The proposed nucleus  $^7$Li is, with percentage corrections, basically a polarized proton in a nuclear medium. In some sense the experiment reminds us of the  the proton spin  analysis \cite{Anselmino:1994gn}, but here with its parton distributions modified by nuclear dynamics. 


%

Our calculation has shown two important results. The most relevant one is that a dramatic change in the EMC region in $R_{pol}$ would imply a dramatic change in the value of the scaling parameters $\eta$ and $\delta$ and therefore an important influence of the spin-spin interactions in nuclei which lie dormant in the average procedure of the unpolarized EMC effect. The second result, which has been confirmed by the MSS approach is that the sea contribution is important for low $x$ if the ratio is performed against the free proton $g_1$ structure function. If one would perform the ratio with respect to the Deuteron, with its pionic sea included, the sea effect would be quantitatively diminished.
Therefore, our study strongly supports the realization of the proposed experiment \cite{Joo} in the whole range of the variable $x$  as the means of better understanding the structure of the nucleon parton distributions in nuclei and their relation to nuclear dynamics.

\section*{Acknowledgements}
We acknowledge Elliot Leader, who proposed the idea of the present analysis and for his careful reading of the manuscript.
We thank Rodolfo Sassot and  Robert Thorne for invaluable assistance in the use of their codes, and Arcadi Santamaria for  
sharing his ample knowledge of Mathematica.   We have all been partially supported  ANPCyT Argentina. V.V. has been also supported by the Ministerio de Econom\'ia y Competitividad and  EU FEDER under contract FPA2010-21750-C02-01, by Generalitat Valenciana: Prometeo/2009/129 and by the EPLANET network under contract PIRSES-2009-GA-246806.


\begin{thebibliography}{ABC}
\section*{References}


\bibitem{Aubert:1983xm}
  J.~J.~Aubert {\it et al.}  [European Muon Collaboration],
  Phys.\ Lett.\ B {\bf 123} (1983) 275.

\bibitem{Rith:2014tma}
  K.~Rith,
  arXiv:1402.5000 [hep-ex].

\bibitem{Higinbotham:2010tb}
  D.~W.~Higinbotham,
  AIP Conf.\ Proc.\  {\bf 1374}, (2011) 85.
  [arXiv:1010.4433 [nucl-ex]].

\bibitem{Weinstein:2010rt}    
  L.~B.~Weinstein, E.~Piasetzky, D.~W.~Higinbotham, J.~Gomez, O.~Hen and R.~Shneor,
  Phys.\ Rev.\ Lett.\  {\bf 106}, (2011) 052301.
  [arXiv:1009.5666 [hep-ph]].

\bibitem{Piasetzky:2011zz}
  E.~Piasetzky, L.~B.~Weinstein, D.~W.~Higinbotham, J.~Gomez, O.~Hen and R.~Shneor,
  Nucl.\ Phys.\ A {\bf 855}, (2011) 245.
  
\bibitem{Arrington:2012ax}
  J.~Arrington, A.~Daniel, D.~Day, N.~Fomin, D.~Gaskell and P.~Solvignon,
  Phys.\ Rev.\ C {\bf 86} (2012) 065204
  [arXiv:1206.6343 [nucl-ex]].
  
\bibitem{Hen:2012fm} 
  O.~Hen, E.~Piasetzky and L.~B.~Weinstein,
  Phys.\ Rev.\ C {\bf 85}, (2012) 047301.
  [arXiv:1202.3452 [nucl-ex]].
  
\bibitem{Frankfurt:2009vv}
  L.~Frankfurt, M.~Sargsyan and M.~Strikman,
  AIP Conf.\ Proc.\  {\bf 1056}, (2008) 322.
  [arXiv:0901.2340 [nucl-th]].

  
\bibitem{Vanhalst:2012ur}
  M.~Vanhalst, J.~Ryckebusch and W.~Cosyn,
  Phys.\ Rev.\ C {\bf 86} (2012) 044619
  [arXiv:1206.5151 [nucl-th]].
  


\bibitem{GarciaCanal:2013dma}
  C.~A.~Garca Canal, T.~Tarutina and V.~Vento,
  Eur.\ Phys.\ J.\ A {\bf 49}, 105 (2013).

\bibitem{Joo} K. Joo et al.: The EMC Effect in Spin Structure Functions, A 12 GeV letter of intent to Jefferson Lab PAC 35 (2009)

\bibitem{Cloet:2005rt}
  I.~C.~Cloet, W.~Bentz and A.~W.~Thomas,
  Phys.\ Rev.\ Lett.\  {\bf 95} (2005) 052302
  [nucl-th/0504019].
  
\bibitem{Cloet:2006bq}
  I.~C.~Cloet, W.~Bentz and A.~W.~Thomas,
  Phys.\ Lett.\ B {\bf 642} (2006) 210
  [nucl-th/0605061].




\bibitem{Smith:2005ra}
  J.~R.~Smith and G.~A.~Miller,
  Phys.\ Rev.\ C {\bf 72} (2005) 022203
  [nucl-th/0505048].



\bibitem{Ganesamurthy:2011zza}
  K.~Ganesamurthy and R.~Sambasivam,
  Nucl.\ Phys.\ A {\bf 856} (2011) 112.

\bibitem{Frankfurt:1981mk}
  L.~L.~Frankfurt and M.~I.~Strikman,
  Phys.\ Rept.\  {\bf 76} (1981) 215.
  
\bibitem{Jaffe:1988up}
  R.~L.~Jaffe and A.~Manohar,
  Nucl.\ Phys.\ B {\bf 321} (1989) 343.



\bibitem{PAC29}  JLAB PAC29 Report on Letter of Intent: LOI-06-004.


\bibitem{GarciaCanal:1984eh}
  C.~A.~Garcia Canal, E.~M.~Santangelo and H.~Vucetich,
  Phys.\ Rev.\ Lett.\  {\bf 53}, 1430 (1984).

\bibitem{Epele:1994aq} 
  L.~N.~Epele, H.~Fanchiotti, C.~A.~Garcia Canal, R.~Sassot and E.~Leader,
  Z.\ Phys.\ C {\bf 64}, 285 (1994).


  
\bibitem{deFlorian:1993hx}
  D.~de Florian, L.~N.~Epele, H.~Fanchiotti, C.~A.~Garcia Canal and R.~Sassot,
  Z.\ Phys.\ A {\bf 350} (1994) 55.


\bibitem{GarciaCanal:1986xe}
  C.~A.~Garcia Canal, E.~M.~Santangelo and H.~Vucetich,
  Phys.\ Rev.\ D {\bf 35}, 382 (1987).



\bibitem{Close:1984zn}
  F.~E.~Close, R.~L.~Jaffe, R.~G.~Roberts and G.~G.~Ross,
  Phys.\ Rev.\ D {\bf 31} (1985) 1004.



\bibitem{Martin:2009iq}
  A.~D.~Martin, W.~J.~Stirling, R.~S.~Thorne and G.~Watt,
  Eur.\ Phys.\ J.\ C {\bf 63} (2009) 189
  [arXiv:0901.0002 [hep-ph]].



\bibitem{deFlorian:2008mr}
  D.~de Florian, R.~Sassot, M.~Stratmann and W.~Vogelsang,
  Phys.\ Rev.\ Lett.\  {\bf 101} (2008) 072001
  [arXiv:0804.0422 [hep-ph]].

\bibitem{deFlorian:2009vb}
  D.~de Florian, R.~Sassot, M.~Stratmann and W.~Vogelsang,
  Phys.\ Rev.\ D {\bf 80} (2009) 034030
  [arXiv:0904.3821 [hep-ph]].



\bibitem{Pieper:2004qw}
  S.~C.~Pieper, R.~B.~Wiringa and J.~Carlson,
  Phys.\ Rev.\ C {\bf 70} (2004) 054325
  [nucl-th/0409012].


\bibitem{Walliser:1985zz}
  H.~Walliser and T.~Fliessbach,
  Phys.\ Rev.\ C {\bf 31} (1985) 2242.



\bibitem{Anselmino:1994gn}
  M.~Anselmino, A.~Efremov and E.~Leader,
  Phys.\ Rept.\  {\bf 261} (1995) 1
   [Erratum-ibid.\  {\bf 281} (1997) 399]
  [hep-ph/9501369].


\end{thebibliography}
\end{document}